\documentclass[reprint,amssymb, amsmath, aps, showpacs, footinbib, prb]{revtex4-1}
\usepackage{graphicx,epstopdf} %Include figure files,superscriptaddress
\usepackage{amsmath}
\usepackage{tabu}
\usepackage{gensymb}
\usepackage{booktabs}
\usepackage{tabularx}

\newcommand{\be}{\begin{equation}}
\newcommand{\ee}{\end{equation}}
\newcommand{\bea}{\begin{eqnarray}}
\newcommand{\eea}{\end{eqnarray}}
\newcommand{\bse}{\begin{subequations}}
	\newcommand{\ese}{\end{subequations}}

\begin{document}
\title{Cluster spin-glass behaviour and memory effect in Cr$_{0.5}$Fe$_{0.5}$Ga}
\author{Pallab Bag}
\author{P. R. Baral}
\author{R. Nath}
\email{rnath@iisertvm.ac.in}
\address{School of Physics, Indian Institute of Science Education and Research Thiruvananthapuram, Kerala-695016, India}
\date{\today}

\begin{abstract}
We report the structural, static, and dynamic properties of Cr$_{0.5}$Fe$_{0.5}$Ga by means of powder x-ray diffraction, DC magnetization, heat capacity, AC susceptibility, magnetic relaxation, and magnetic memory effect measurements. DC magnetization and AC susceptibility studies reveal a spin-glass transition at around $T_{\rm f} \simeq 22$~K. An intermediate value of the relative shift in freezing temperature $\delta T_{\rm f} \simeq 0.017$, obtained from the AC susceptibility data reflects the formation of cluster spin-glass states. The frequency dependence of $T_{\rm f}$ is also analyzed within the framework of dynamic scaling laws such as power law and Vogel-Fulcher law. The analysis using power law yields a characteristic time constant for a single spin flip $\tau* \simeq 1.1\times10^{-10}$~s and critical exponent $z\nu^{\prime}=4.2\pm0.2$. On the other hand, the Vogel-Fulcher law yields the characteristic time constant for a single spin flip $\tau_0 \simeq 6.6\times10^{-9}$~s, Vogel-Fulcher temperature $T_{\rm 0}=21.1\pm0.1$~K, and an activation energy $E_{\rm a}/k_{\rm B} \simeq 16$~K. The value of $\tau*$ and $\tau_0$ along with a non-zero value of $T_{\rm 0}$ provide further evidence for the cluster spin-glass behaviour. The magnetic field dependent $T_{\rm f}$ follows the de Almeida-Thouless (AT) line with a non-mean-field type instability, reflecting either a different universality class or strong anisotropy in the spin system. A detailed non-equilibrium dynamics study via relaxation and memory effect experiments demonstrates the evolution of the system through a number of intermediate metastable states and striking memory effects. All the above observations render a cluster spin-glass behaviour in Cr$_{0.5}$Fe$_{0.5}$Ga which is triggered by magnetic frustration due to competing antiferromagnetic and ferromagnetic interactions and magnetic site disorder. Moreover, the asymmetric response of magnetic relaxation with respect to the change in temperature, below the freezing temperature can be explained by the hierarchical model.
\end{abstract}
%\keywards{Fe-based intermetallic, spin-glass, memory effect}
\maketitle

\section{Introduction}
In the past years, alloys showing spin-glass (SG) behaviour have been widely pursued in order to study exchange bias effect, slow dynamics, memory effect, aging effect etc.\cite{Fisher1991Book,Mydosh1993Book,Binder1986RMP,Weissman1993RMP,Jonason1998PRL,Fisher1988PRB,Karmakar2008PRB,Chatterjee2009PRB}
%Such materials are relevant for the purpose of ultrahigh-density magnetic recording and spin electronic devices.\cite{Karmakar2008PRB,Wang2004PRB,Grover1988PRB,Lue2001PRB,Jonason1998PRL}
SG is basically a disordered ground state where the spins are frozen along arbitrary direction, below a critical temperature. It is commonly believed that SG appears in systems where magnetic long-range-ordering (LRO) is disturbed by site disorder and magnetic frustration.\cite{Binder1986RMP,Mydosh1993Book} Examples of such systems include metallic SGs where magnetic impurities are randomly diluted in a nobel metal, geometrically frustrated lattices where lattice topology precludes the minimization of energy, systems frustrated due to competing antiferromagnetic (AFM) and ferromagnetic (FM) interactions or competing nearest neighbour and next nearest neighbour interactions etc.\cite{Souletie1978JP,Gardner2010RMP} SG-like non-equilibrium dynamics has been observed in several systems where the basic building blocks responsible for the glassy behaviour are bigger spin entities, rather than individual spins, referred as "spin-clusters". Such systems are often characterized by slow dynamics, similar to the classical SGs.
%Magnetic frustration is of two types: (i) geometrical frustration which arises due to the geometry of the spin-lattice and (ii) frustration due to completing interactions.
%Diluted magnetic materials often show SG behaviour since chemical substitution induces either site disorder or competing interactions. 
Despite an extensive study on SGs, a consensus about the ground state and dynamics in these systems is still lacking.
%In certain cases, the low temperature disordered (SG) state is preceded by a long-range antiferromagnetic (AFM) or ferromagnetic (FM) ordering, such materials are know as reentrant spin-glass (RSG).\cite{Binder1986RMP,Mydosh1993Book,Abiko2005RRL,Verbeek1978PRL,Chatterjee2009PRB}
Here, we report the magnetic studies on the diluted alloy Cr$_{0.5}$Fe$_{0.5}$Ga which exhibits features that are reminiscent of cluster SG.

The iso-structural alloys CrGa and FeGa crystallize in a Cr$_5$Al$_8$-type rhombohedral structure (space group $R\bar{3}m$) with lattice constants [$ a = 12.625 (8) $~\text{\AA} and $ c = 7.785 (10) $~\text{\AA}] and [$ a = 12.4368 (11) $~\text{\AA} and $ c = 7.7642 (10) $~\text{\AA}], respectively.\cite{Gourdon2004IC} In the unit cell, both Ga and Cr/Fe occupy three inequivalent sites each. They form two types of icosahedra: one is Ga-centered and the other one is Cr/Fe-centered which are alternating along the crystallographic $c$-direction forming chains. Magnetic susceptibility of CrGa is almost temperature independent while for FeGa, it shows a peak at $\sim42$~K and a broad maximum at $\sim$135~K. Band structure calculations predict weak AFM and dominant FM exchange couplings for CrGa and FeGa, respectively.\cite{Gourdon2004IC}
%Indeed, a Curie-Weiss fit to the high temperature susceptibility data of FeGa yields a large and positive value of Curie-Weiss temperature $\theta_{\rm CW} \simeq 134$~K, reflecting that the dominant interaction is FM.
%The ground state of CrGa is reported to be non-magnetic or weakly AFM while for FeGa, the ground state is FM.\cite{Gourdon2004IC}
Therefore, substitution of Fe at the Cr site in CrGa can alter the AFM interaction among the Cr atoms and induce different magnetic states. Ko $ et~al $ tried to synthesize Cr$_{1-x}$Fe$_x$Ga for different values of $x$ but they succeeded to synthesize phase pure sample only for $x=0.5$.\cite{Ko2010IC} Neutron powder diffraction on Cr$_{0.5}$Fe$_{0.5}$Ga revealed average composition of the powder sample to be Cr$_{0.515}$Fe$_{0.485}$Ga with lattice constants $ a = 12.5448 (4) $~\text{\AA} and $ c = 7.8557 (2)$~\text{\AA} at room temperature and partial ordering of Cr and Fe atoms among three crystallographic sites. In particular, the Cr and Fe atoms occupy three inequivalent sites: M1(3b), M2(18h), and M3(18h). The refined site occupancies for Cr/Fe are 0.587/0.413, 0.636/0.364, and 0.383/0.617, respectively. Preliminary magnetization measurements suggest the onset of a magnetic ordering at $T\simeq 25$~K. Subsequent theoretical calculations indicated that the Fe-Fe and Cr-Fe interactions are FM and AFM, respectively with an overall ferrimagnetic ordering at low temperature.\cite{Ko2010IC} However, a clear understanding of the ground state properties of this alloy requires a detailed experimental investigation which is not yet done.

In this work, we carried out a comprehensive study of the structural and magnetic properties of Cr$_{0.5}$Fe$_{0.5}$Ga. The 50 \% Fe substitution at the Cr site induces atomic disorder in the lattice, preserving the original crystal structure. It is found to be a magnetically frustrated system which undergoes a SG transition at low temperatures. The DC magnetization along with the AC susceptibility data render the system a cluster SG-type. Finally, the magnetic memory effect in the system has been demonstrated by the magnetic relaxation and memory effect measurements.

\section{Experimental details}	
Polycrystalline Cr$_{0.5}$Fe$_{0.5}$Ga sample was synthesized by the conventional solid state reaction technique, taking the constituent elements in the desired stoichiometry. The elements (Fe, Cr, and Ga) used here are of high pure (99.99\%) obtained from Sigma Aldrich. The stoichiometric amounts were sealed in a quartz tube in Ar atmosphere. The ampoule was first heated at $1050~^0$C for 3~days and then at $850~^{0}$C for 5~days. The powder x-ray diffraction (XRD) measurements were carried out (PANalytical powder diffractometer with Cu$K_\alpha$ radiation) as a function of temperature using a low temperature attachment (Oxford Phenix). DC and AC magnetization ($M$) measurements were performed using a vibrating sample magnetometer (VSM) attachment to the physical property measurement system (PPMS, Quantum Design). Heat capacity ($C_{\rm p}$) was measured using the heat capacity option in the PPMS, adopting the relaxation technique. 

\section{Results and discussion}
\subsection{X-ray Diffraction}
\begin{figure}[t!]
	\centering
	\includegraphics[width=\linewidth,height=\linewidth]{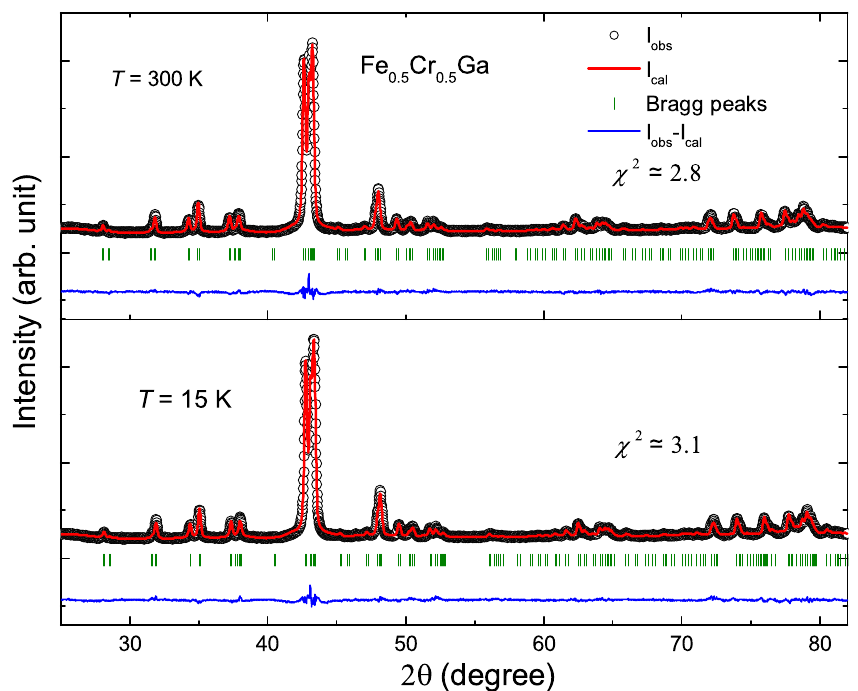}
	\caption{Rietveld refinement of the x-ray diffraction pattern of Cr$_{0.5}$Fe$_{0.5}$Ga at $T=300$~K (upper panel) and $T=15$~K (lower panel), respectively. The open circles and solid lines are the observed and calculated patterns, respectively. The Bragg positions are indicated by ticks. Solid line at the bottom represents the difference between the observed and calculated intensities.}
	\label{Fig1}
\end{figure}

\begin{figure}[t!]
	\centering
	\includegraphics[width=\linewidth,height=\linewidth]{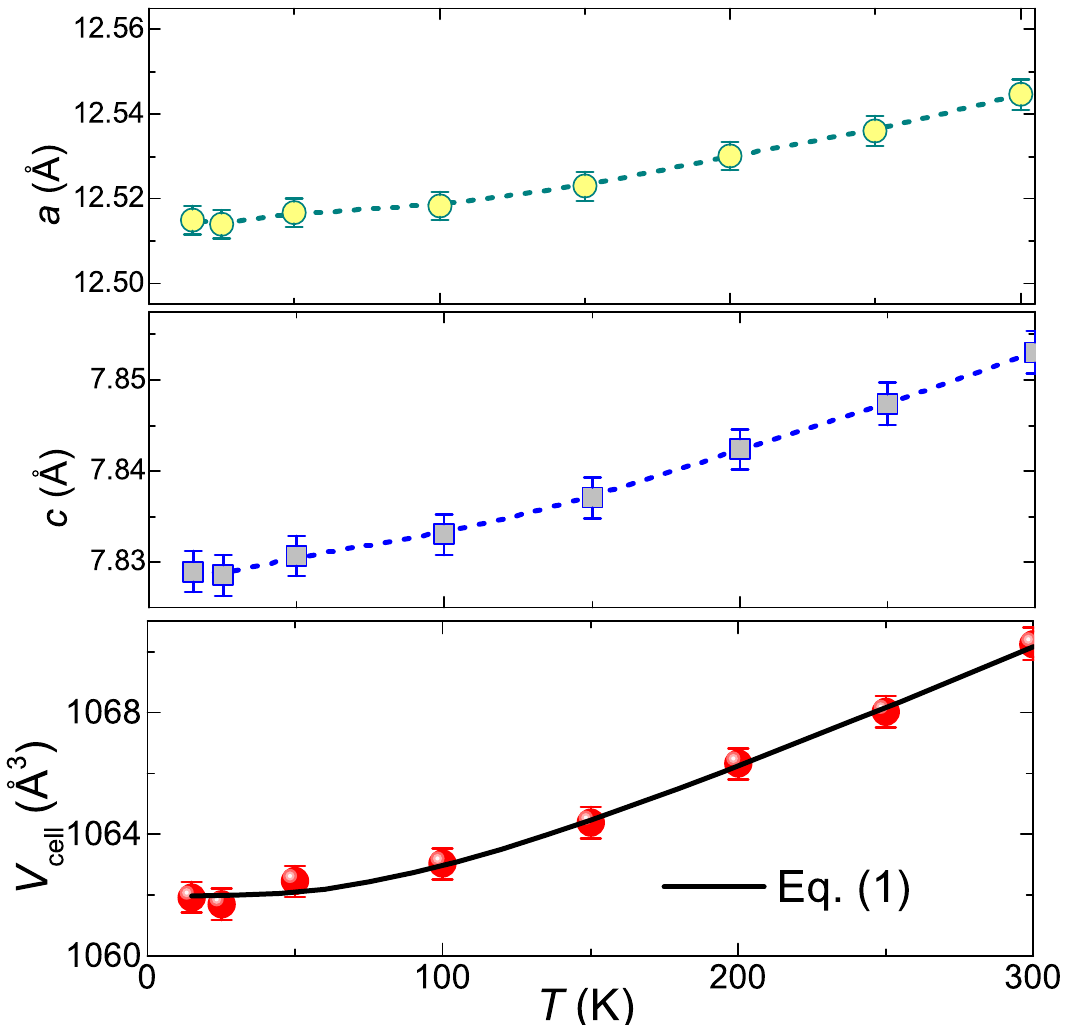}
	\caption{Variation of lattice constants ($a$ and $b$) and unit cell volume ($V_{\rm cell}$) with temperature for Cr$_{0.5}$Fe$_{0.5}$Ga. The solid line represents the fit of $V_{\rm cell}(T)$ using Eq.~(\ref{VcellvsT}).}
	\label{Fig2}
\end{figure}
In order to confirm the phase purity and to detect the structural transition, if any, powder XRD was measured at different temperatures. Rietveld refinement of the XRD pattern was carried out using the FULLPROF package.\cite{Rodriguez1993PhysicaB} The initial structural parameters for this purpose were taken from Ref.~[\onlinecite {Gourdon2004IC}]. Figure~\ref{Fig1} shows the Rietveld refinement of the powder XRD pattern at 300~K and 15~K. A good fitting of the room temperature data with a reduced value of goodness-of-fit ($ \chi^{2} \simeq 2.8$) suggests that the sample is phase pure. The obtained lattice constants at room temperature are $ a = 12.544 (4) $~\text{\AA} and $ c = 7.853 (2) $~\text{\AA} which are consistent with the previous report.\cite{Ko2010IC} Figure~\ref{Fig2} displays the temperature variation of lattice constants and unit cell volume ($V_{\rm cell}$). No structural transition was observed down to 15~K and the lattice constants and $V_{\rm cell}$ were found to decrease systematically with decreasing temperature. The temperature variation of $V_{\rm cell}$ was fitted by the equation\cite{Pakhira2016PRB}
\begin{equation}
V(T)=\gamma U(T)/K_0+V_0,
\label{VcellvsT}
\end{equation}
where $V_0$ is the cell volume at $T = 0$~K, $K_0$ is the bulk modulus, and $\gamma$ is the Gr$\ddot{\rm u}$neisen parameter. $U(T)$ is the internal energy which can be expressed in terms of the Debye approximation as,
\begin{equation}
U(T)=9pk_{\rm B}T\left(\frac{T}{\theta_{\rm D}}\right)^3\int_{0}^{\theta_{\rm D}/T}\dfrac{x^3}{e^x-1}dx.
\end{equation}
Here, $p$ is the number of atoms in the specimen and $k_{\rm B}$ is the Boltzmann constant. Using this approximation (see the fit in the lower panel of Fig.~\ref{Fig2}), the Debye temperature ($\theta_{\rm D}$) for Cr$_{0.5}$Fe$_{0.5}$Ga was estimated to be $\theta_{\rm D} \simeq 350$~K.
%\begin{table}[htb!]
%	\setlength{\tabcolsep}{0.32cm}
%	\caption{Fractional atomic coordinates and occupancies of Ga and Fe/Cr in the unit cell of Cr$_5$Al$_8$-type rhomboidal crystal structure at 300~K (first set of lines) and 15~K (second set of lines), obtained from the Rietveld refinement of the powder XRD data.}
%	\label{Table1}
%\begin{tabular} {c c cccc}
%	\hline\hline
%	Atoms & $x$ & $y$ &  $z$ & Occ.  \\
%	\hline	Ga1 & 0 & 0 &  0 & 0.083\\
%	 & 0 & 0 &  0 & 0.83\\
%	Ga2 & 0.2354 (3) & 0.1177 (2) &  0.5738 (4) & 0.5\\
%	 & 0.2350 (3) & 0.1175 (3) &  0.5741 (4) & 0.5\\
%	Ga3 & 0.2789 (3) & -0.0544 (3) &  0.0833 (3) & 0.5\\
%	& 0.2792 (3) & -0.0541 (3) &  0.8333 (4) & 0.5\\
%	Fe1 & 0 & 0 & 0.5 & 0.0416 \\
%	 & 0 & 0 & 0.5 & 0.0416 \\
%	Fe2 & 0.2366 (3) & 0.1183 (3) & 0.9224 (5) & 0.25\\
%	 & 0.2360 (3) & 0.1179 (3) & 0.9207 (5) & 0.25\\
%	Fe3 & -0.1478 (3) & -0.0738 (3) &  0.7481 (5) & 0.25\\
%	 & -0.1474 (3) & -0.0736 (4) &  0.7483 (5) & 0.25\\
%	Cr1 & 0 & 0 & 0.5 & 0.0416 \\
%	 & 0 & 0 & 0.5 & 0.0416  \\
%	Cr2 & 0.2366 (3) & 0.1183 (3) & 0.9224 (3) & 0.25\\
%	 & 0.2360 (3) & 0.1179 (3) & 0.9207 (5) & 0.25\\
%	Cr3 & -0.1478 (4) & -0.0738 (3) &  0.7481 (4) & 0.25\\
%	& -0.1474 (3) & -0.0736 (3) &  0.7483 (5) & 0.25\\
%	\hline\hline
%\end{tabular}
%\end{table}

\subsection{DC Magnetization}
 \begin{figure}[t!]
 	\centering
 	\includegraphics[width=\linewidth,height=0.9\linewidth]{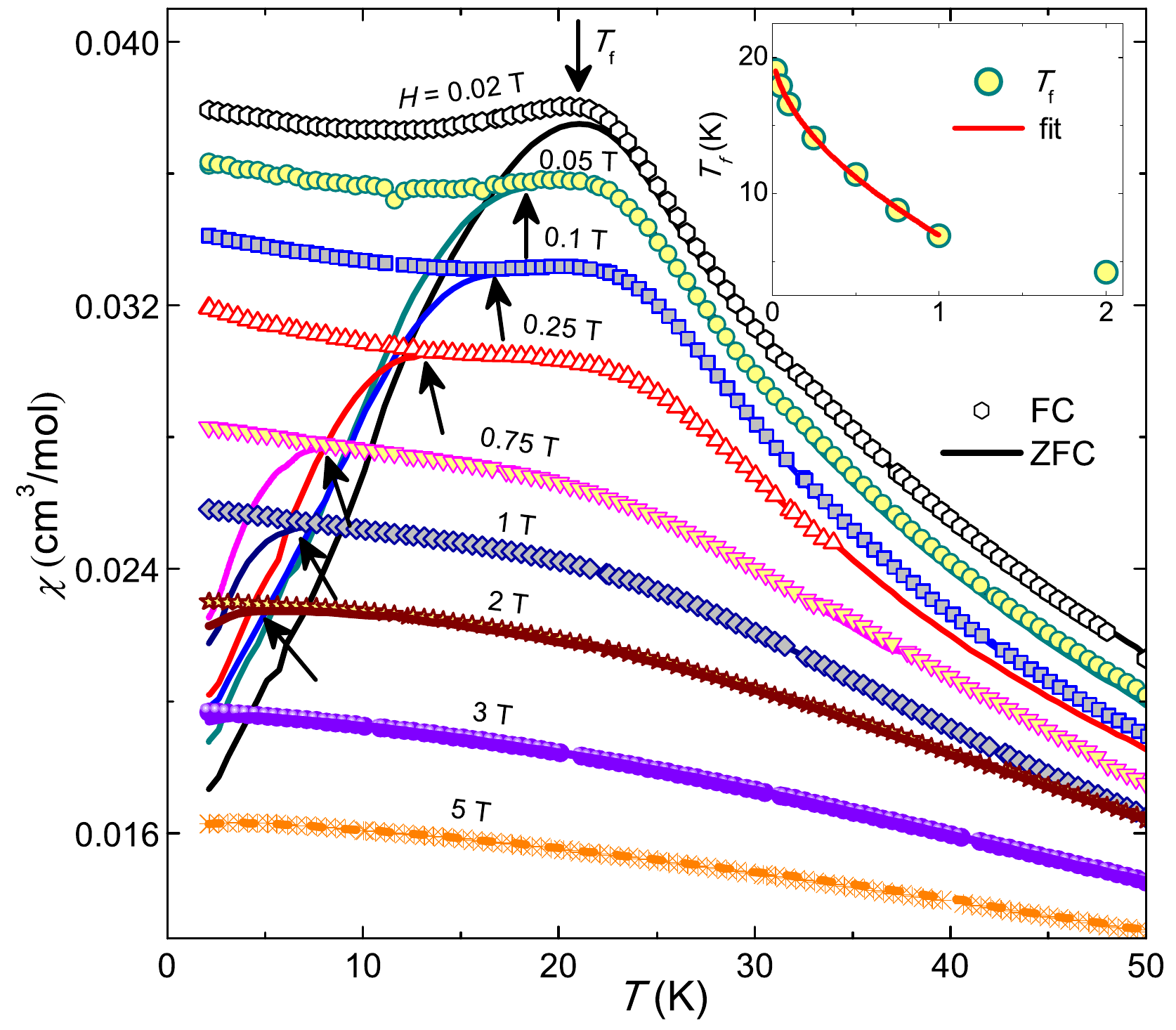}
 	\caption{Temperature dependent DC susceptibility $\chi(T)$ measured under different applied fields for ZFC and FC protocols. The arrows point to $T_{\rm f}$. Inset: the variation of $T_{\rm f}$ with $H$. The solid line represents the fit using Eq.~(\ref{Eq_TfvsH})}
 	\label{Fig3}
 \end{figure}
Figure~\ref{Fig3} presents the temperature dependent DC susceptibility, $\chi(T)~(\equiv M/H)$ measured in different applied fields, during heating after zero-field-cooled (ZFC) and field-cooled (FC) conditions. For $H=0.02$~T, both ZFC and FC data show a broad peak and a bifurcation at the same temperature possibly suggesting a glass transition around 22~K. The temperature at which the bifurcation occurs is called the freezing temperature, denoted as $T_{\rm f}$. In order to elucidate the nature of the transition, we measured $\chi(T)$ at different applied fields for ZFC and FC protocols. As the field increases, the absolute value of $\chi$ decreases systematically and the ZFC data develop a plateau with two broad edges on either side. The low temperature edge corresponding to $T_{\rm f}$ shifts towards lower temperatures while the edge at the high temperature side shifts towards higher temperatures. Furthermore, the difference between ZFC and FC curves ($\Delta \chi$) at low temperatures decreases with increasing magnetic field. The shifting of $T_{\rm f}$ towards lower temperatures and the reduction in $\Delta\chi$ indicate the frozen spin-glass (SG) state below $T_{\rm f}$.\cite{Binder1986RMP} For the field above 3~T, $T_{\rm f}$ is suppressed below 2~K and hence not detectable. The shifting of high temperature edge towards high temperatures with field appears to be due to the onset of a magnetic LRO. However, our heat capacity measurement (discussed later) rules out any magnetic ordering at this temperature. The overall behaviour of $\chi(T)$ is nearly similar to that reported for SG compound U$_2$PdSi$_3$.\cite{Li1998PRB}

\begin{figure}[t!]
	\centering
	\includegraphics[width=\linewidth,height=\linewidth]{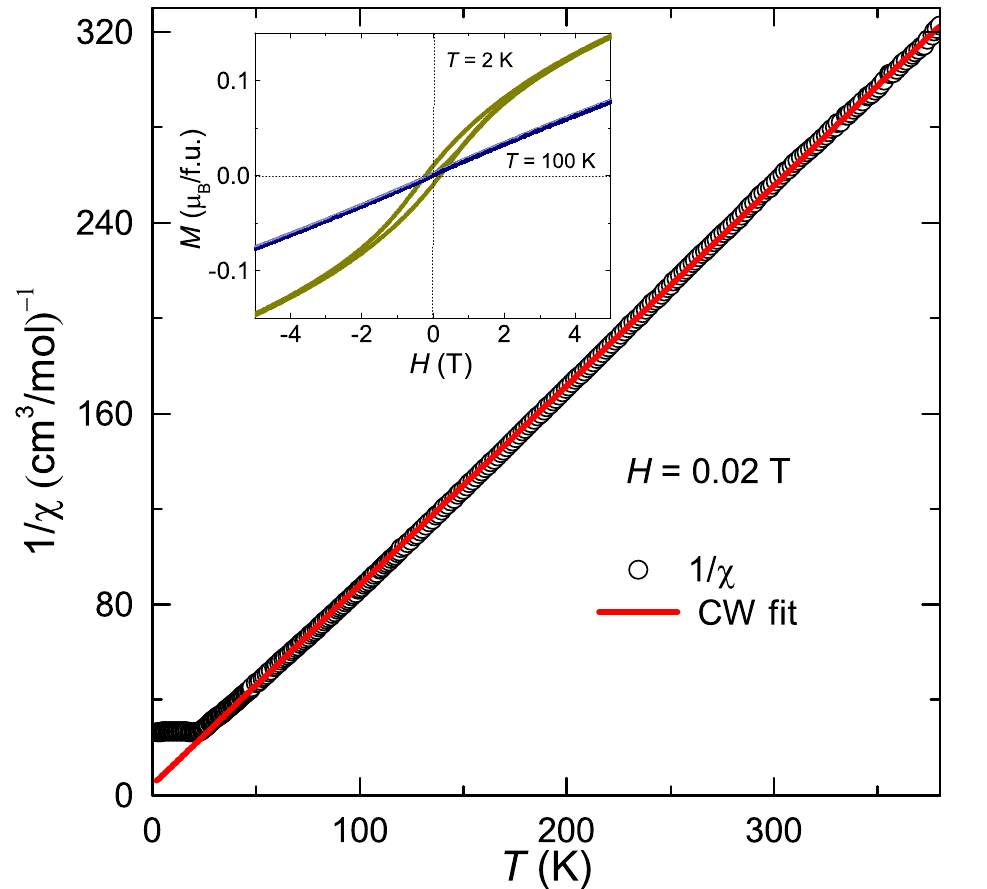}
	\caption{$1/\chi$ vs $T$ measured at $H=0.02$~T and the solid line is the CW fit for $T>50$~K. Inset: Isothermal magnetization $M(H)$ curves at $T=2$~K and 100~K.}
	\label{Fig4}
\end{figure}
Magnetic isotherm [$M(H)$] measurements were also performed at different temperatures (inset of Fig.~\ref{Fig4}). At high temperatures, $M(H)$ is nearly straight line, as expected in the paramagnetic (PM) region. With decreasing temperature, it develops a curvature which is more pronounced at low temperatures. At the lowest measured temperature of $T=2$~K, it shows a small hysteresis with a coercive field of $\sim 200$~Oe.
However, the value of magnetization even at 9~T (not shown) is much less than the saturation value expected for this alloy. A weak hysteresis and the reduced value of magnetization at 9~T exclude the possibility of a FM/ferrimagnetic transition and establishes low temperature SG behaviour of the compound.\cite{Binder1986RMP,Malinowski2011PRB} 

%This is of course in accordance with the theoretical predictions by Ko et al.\cite{Ko2010IC} Thus, the SG transition is preceded by a ferrimagnetic LRO at a slightly higher temperature which is reminiscent of the occurrence of a reentrant spin-glass (RSG) transition. Such type of behaviour is well documented in some of the previously reported compounds.\cite{Binder1986RMP,Abiko2005RRL,Pakhira2016PRB,maji2011JPCM,Luo2015JAP,Ding2015JAP,Ghara2014PRB,Verbeek1978PRL,Chatterjee2009PRB}

As shown in Fig.~\ref{Fig4}, the inverse susceptibility $1/\chi$ (measured at $H=0.02$~T) in the high temperature regime ($T>50$~K) is fitted by the Curie-Wiess (CW) law 
\begin{equation}
\chi = \frac{C}{T - \theta_{\rm{CW}}},
\end{equation}
where $ C $ and $ \theta_{\rm{CW}} $ represent the Curie constant and CW temperature, respectively. The obtained values are $C\simeq1.2~$cm$^3$K/mol and $ \theta_{\rm{CW}}\simeq-5$~K. The negative value of $ \theta_{\rm{CW}}$ signifies the presence of dominant AFM interaction in the system. From the value of $C$, the effective magnetic moment $\mu_{\rm eff}$ (= $\sqrt{3k_{\rm{B}}C/N_{\rm{A}}}$, where $N_{\rm A}$ is the Avogadro's number) was calculated to be $\sim 3.1 \mu_{\rm{B}} $.
%$ \mu_{\mathrm{eff}} =\sqrt{0.5\mu_{\mathrm{Fe}}^2+0.5\mu_{\mathrm{Cr}}^2}=4.2 \mu_{\mathrm{B}}$ for this alloy.\cite{Ko2010IC}
In spin systems, according to the mean-field theory $\theta_{\rm {CW}}$ represents the sum of all the exchange couplings. Our estimated value of $\theta_{\rm {CW}}$ is much smaller compared to $T_{\rm f}$, which possibly reflects that the system is frustrated due to competing AFM and FM interactions,\cite{Nath2008PRB} as anticipated from the previous theoretical calculations.\cite{Ko2010IC} In addition to the magnetic site disorder, this competing interactions is also responsible for the low temperature SG behaviour of the system.

%and one can quantify frustration in the system using the parameter %$f=\frac{\mid\theta_{\rm{CW}}\mid}{T_{\rm N}}$, where $T_{\rm N}$ is the magnetic %long-range-ordering (LRO). However, the system under investigation doesnot undergo any %magnetic-LRO. As an approximation, taking the value of $T_f$ as the upper limit for %$T_{\rm N}$, we get $f\simeq 0.24$. For a frustrated AFM, typically, one observes a very %large value of $f$ ($>1$). Interestingly, our estimated value of $\theta_{\rm {CW}}$ is %much smaller compared to the transition temperature, giving raise to a reduced value of %$f$ ($<1$). This reflects that the system is highly frustrated due to competing AFM and %FM interactions,\cite{Nath2008PRB} as anticipated from the previous theoretical %calculations.\cite{Ko2010IC} In addition to the magnetic site disorder, this competing %interactions is also responsible for the low temperature SG behaviour of the system.

%Similar $T_{\rm f}$ vs $H$ behaviour has been reported earlier for quite a few other SG systems.\cite{Ghara2014PRB,Pakhira2016PRB,Luo2015JAP,Lago2012PRB}

The variation of $T_{\rm f}$ with $H$ in the low field region is presented in the inset of Fig.~\ref{Fig3}. It decreases systematically with increasing field, consistent with the SG transition. In the $H-T$ phase diagram for SG systems, typically, two irreversible lines are observed: Gabay-Toulouse (GT) line ($T_f \propto H^2$) and de Almeida-Thouless (AT) line.\cite{Gabay1981PRL,Ameida1978JPA} The AT line marks the PM to SG transition which is usually observed for Ising spin systems. On the other hand, in the case of Heisenberg spin systems, both the lines are expected. In the strong anisotropy (strong irreversibility) regime, the system is Ising-type and the line follows AT character whereas in the weak anisotropy regime, the line corresponds to GT line.
A quantitative difference is expected in the behavior of AT line in the mean-field and non-mean-field scenarios. According to the non-mean-field scaling theory, the variation of $T_{\rm f}$ with $H$ in the low field region follows\cite{Malozemoff1983PRL}
\begin{equation}
T_{\rm f}(H)=T_{\rm f}(0)(1-AH^{2/\Phi}),
\label{Eq_TfvsH}
\end{equation}
where $A$ is the amplitude, $T_{\rm f}(0)$ is the value of $T_{\rm f}$ in the absence of a magnetic field, and $\Phi$ is the crossover exponent. In the mean-field model it has a value $\Phi = 3$. In our system, only one irreversible line was observed which could be fitted by Eq.~(\ref{Eq_TfvsH}). As shown in the inset of Fig.~\ref{Fig3}, the best fit was obtained for $H<1$~T with $T_{\rm f}(0)\simeq 20.7$~K and $\Phi \simeq 3.8$. This value of $\Phi$ is larger than the one expected for the AT line with mean-field instability.\cite{Ameida1978JPA} In several cluster SG systems such as Nd$_2$AgIn$_3$(Ref.~\onlinecite{Li2001APL}), U$_2$IrSi$_3$ (Ref.~\onlinecite{Li2003PRB}), Zn$_3$V$_3$O$_8$ (Ref.~\onlinecite{Chakrabarty2014JPCM}), Nd$_5$Ge$_3$ (Ref.~\onlinecite{Maji2011JPCM}), LiMn$_2$O$_4$ (Ref.~\onlinecite{Zhang2018PRB}) etc, the authors have reported a mean-field type behaviour.\footnote{In these experimental papers the authors have force-fitted their data to a mean-field model rather than attempting a more general non-mean-field type model.} An analysis (reported in Ref.~\onlinecite{Barbara1981PRL,Dieny1986PRL}) in terms of non-mean-field model [Eq.~(\ref{Eq_TfvsH})] on a group of SG systems resulted a large variation of $\Phi$ from 5 for canonical Heisenberg SG system MnCu to 3.2 for random-anisotropy SG system $a$-DyNi, suggesting that they all donot belong to the same universality class. This model has also been tested on other SG systems which produce a non-mean-field type exponent.\cite{Tabata2017PRB,Shand2010PRB,Wang2004PRB} The obtained value of $\Phi \simeq 3.8$ for our system falls in the intermediate range, reflecting either a different universality class or role of dominant anisotropy in the system.
 
%In our system, only one irreversible line was observed and the plot of $T_{\rm f}$ vs $H^{2/3}$ fits satisfactorily with the AT relation, $T_{\rm f} \propto H^{2/3}$ for $H \leq 1$~T. This is an indication of the Ising-type SG state in Cr$_{0.5}$Fe$_{0.5}$Ga. Similar behaviour has been reported earlier for few cluster SG systems such as Nd$_2$AgIn$_3$(Ref.~\onlinecite{Li2001APL}), U$_2$IrSi$_3$ (Ref.~\onlinecite{Li2003PRB}), Zn$_3$V$_3$O$_8$ (Ref.~\onlinecite{Chakrabarty2014JPCM}), Nd$_5$Ge$_3$ (Ref.~\onlinecite{Maji2011JPCM}), LiMn$_2$O$_4$ (Ref.~\onlinecite{Zhang2018PRB}), Fe$_2$O$_3$ (Ref.~\onlinecite{Mukadam2005PRB}) etc.

\begin{figure}[t!]
	\centering
	\includegraphics[width=\linewidth,height=\linewidth]{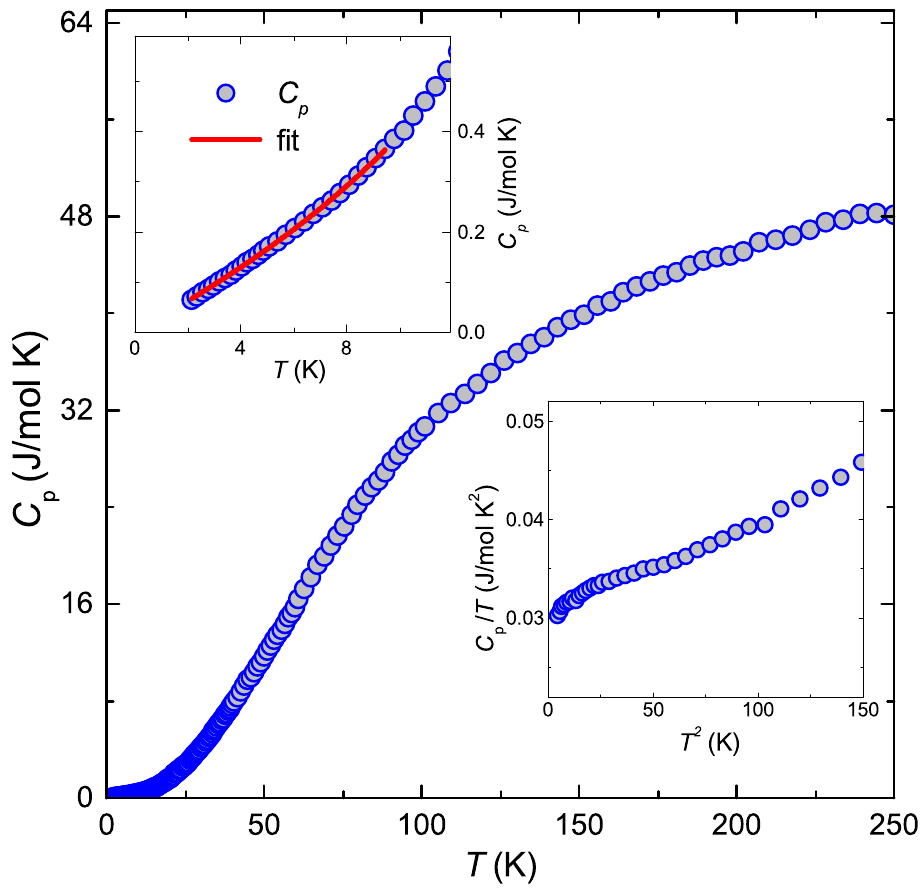}
	\caption{Temperature dependent heat capacity $C_{\rm p}$ in the absence of magnetic field between 2~K and 250~K. Lower inset: $C_{\rm p}/T$ vs $T^2$. Upper inset: $C_{\rm p}$ vs $T$ in the low temperatures regime and solid line is the fit in the temperature range 2 to 10~K, as described in the text.}
	\label{Fig5}
\end{figure} 
\subsection{Heat Capacity}
Figure~\ref{Fig5} shows the temperature dependent $C_{\rm p}$ in the absence of magnetic field. No anomaly associated with the magnetic LRO was observed down to 2~K.
%The non-appearance of any feature at $T_{\rm C} \simeq 22$~K could be due to a very small change in magnetic entropy involved at the weak ferrimagnetic LRO and/or the ordering temperature is high enough that the large phonon contribution masks the associated peak.
The value of $C_{\rm p}$ at $T = 250$~K is about 48.2~J/mol-K which is close to the expected Dulong-Petit value $C_{\rm v}=3mR=6R=49.8$~J/mol~K, where $R$ is the gas constant and $m$ is the number of atoms per formula unit.
In an attempt to check whether one can fit the data in the low temperature region by $C_{\rm p}(T) = \gamma T+ \beta T^3$, $C_{\rm p}/T$ vs $T^2$ is plotted in the lower inset of Fig.~\ref{Fig5}. Here, $\gamma$ is the Sommerfeld coefficient which represents the electronic contribution and $\beta$ represents the lattice contribution. It clearly shows a non-linear behaviour. However, the low temperature $C_{\rm p}$ data could be fitted well by adding a magnetic term $\delta T^{3/2}$ in $C_{\rm p} = \gamma T + \beta T^3$ i.e. $C_{\rm p} = \gamma T + \beta T^3 +\delta T^{3/2}$ where $\delta$ is the co-efficient of $T^{3/2}$.\cite{Anand2012PRB} A $T^{3/2}$ term in $C_{\rm p}$ is typical for SG and FM systems.\cite{Gopal2012Book} The best fit of the data in the temperature range $2-10$~K (upper inset of Fig.~\ref{Fig5}) yields $\gamma \simeq $~29~mJ/mol~K$^2$, $\beta \simeq$~0.072~mJ/mol~K$^4$, and $\delta \simeq 0.7$~mJ/mol~K$^{5/2}$. From the values of $\beta$, one can calculate the Debye temperature ($\theta_{\rm D}$) using the standard expression $\theta_{\rm D} = (12\pi^4 mR/5\beta)^{1/3}$. The value of $\theta_{\rm D}$ is calculated to be $\sim 377$~K which is close to the value obtained from the $V_{\rm cell}$ vs $T$ analysis. A large value of $\gamma$ is reported for several cluster SG systems but the effect of disorder on the density of states is not yet understood.\cite{Anand2012PRB,Li1998PRB,Tien2000PRB}

\subsection{AC Susceptibility}
\begin{figure}[t!]
	\centering
	\includegraphics[width=\linewidth,height=1.4\linewidth]{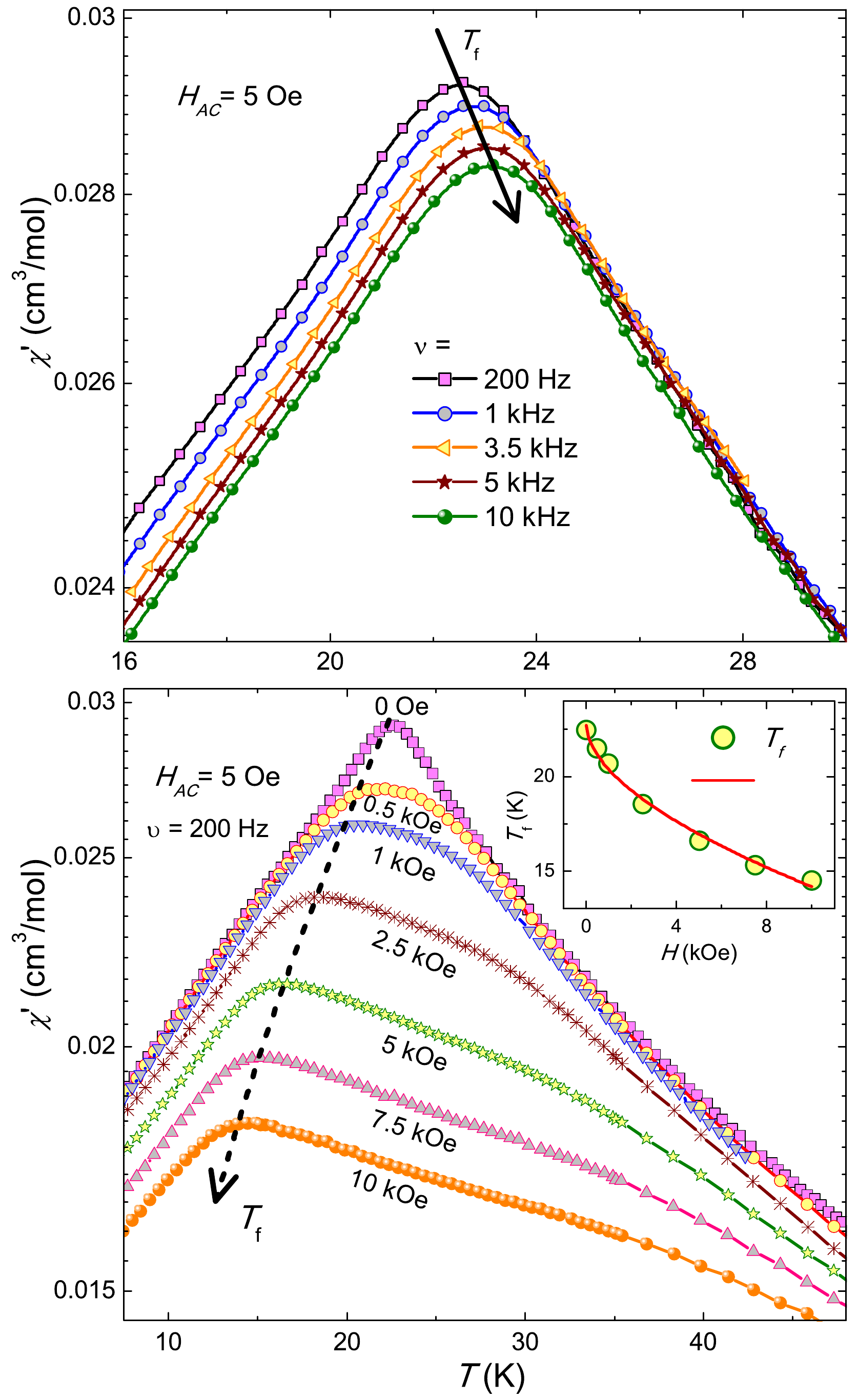}
	\caption{Real part of the AC susceptibility [$\chi'(T)$] measured at different frequencies ($\nu$) and at an AC field $H_{\rm AC}=5$~Oe. The solid downward arrow points to the peak shift. Lower panel: $\chi'(T)$ measured at different DC applied fields, fixing $\nu = 200$~Hz and $H_{\rm AC} = 5$~Oe. The dotted downward arrow guides the peak position. Inset: the variation of $T_{\rm f}$ with $H$. The solid line represents the fit using Eq.~(\ref{Eq_TfvsH}).}
	\label{Fig6}
\end{figure}
In order to understand the underlying nature of the transition and to study the dynamics of the SG state, AC susceptibility was measured at different frequencies ($\nu$) and at a fixed excitation field of $H_{\rm AC}=5$~Oe, after cooling the sample in zero field. The real part of the AC susceptibility ($\chi'$) as a function of temperature is plotted in the upper panel of Fig.~\ref{Fig6}. It exhibits a pronounced anomaly at around $22.5$~K (for $\nu = 200$~Hz) which is found to be frequency dependent. The peak position shifts towards higher temperatures and the height of the peak decreases with increasing $\nu$, consistent with a glassy transition with freezing temperature $T_{\rm f}\simeq 22.5$~K. AC susceptibility was also measured under different DC fields ($H_{DC}$) fixing the AC excitation at $H_{\rm AC} = 5$~Oe and $\nu = 200$~Hz. As one can see in the lower panel of Fig.~\ref{Fig6}, the peak at $T_{\rm f}\simeq22.5$~K in zero-field transforms into a broad shoulder like shape when $H_{\rm DC}$ is applied, similar to the DC susceptibility data. With increasing $H_{\rm DC}$, the low temperature edge ($T_{\rm f}$) moves towards low temperatures, further supporting the SG transition.
The variation of $T_{\rm f}$ with $H$ could also be fitted well using Eq.~\eqref{Eq_TfvsH} (see the inset of Fig.~\ref{Fig6}) which yields $T_{\rm f} (0)\simeq 22.7$~K and $\Phi \simeq 3.6$. The obtained value of $\Phi$ again reflects the de Almeida-Thouless line with non-mean-field instability.

\begin{figure}[t!]
	\centering
	\includegraphics[width=\linewidth,height=1.2\linewidth]{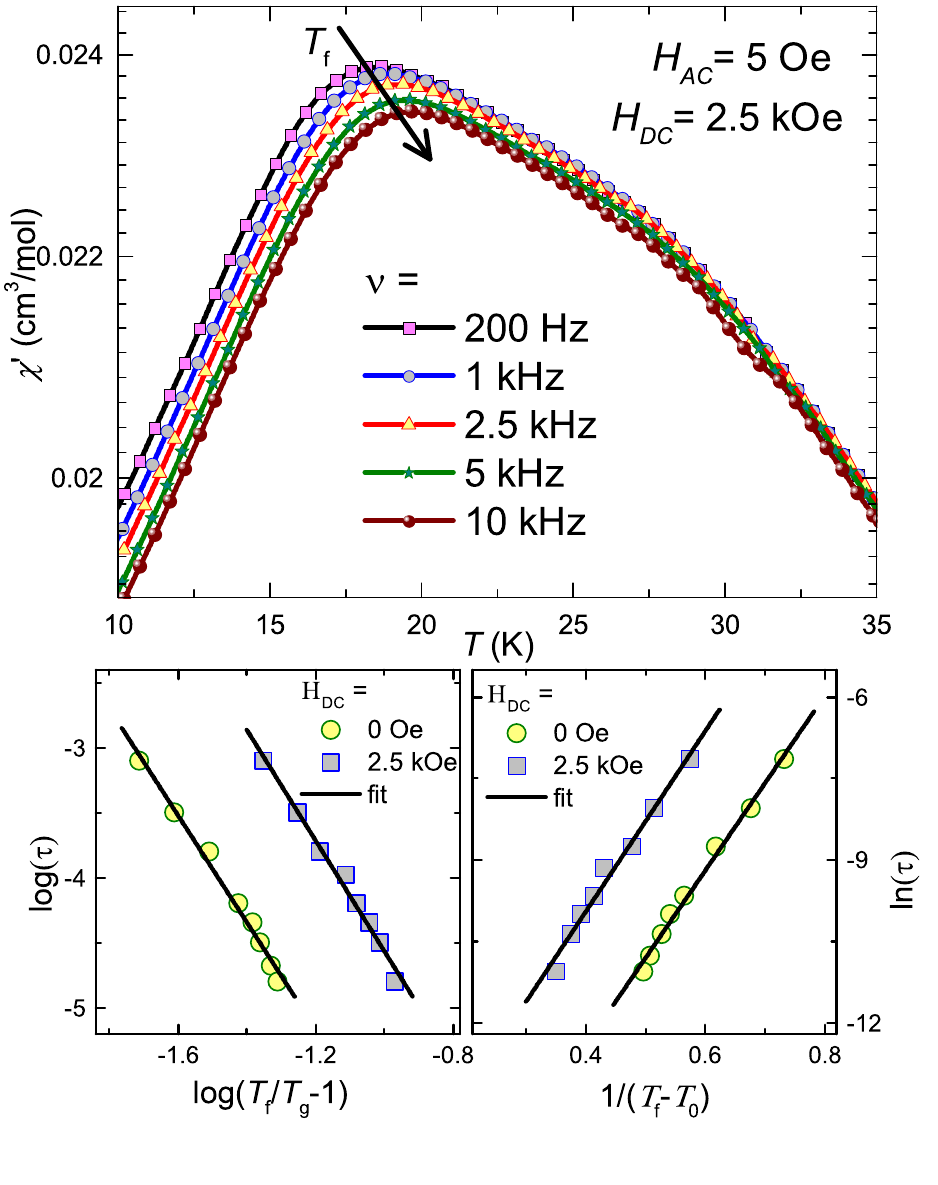}
	\caption{Upper panel: $\chi'(T)$ measured at different frequencies ($\nu$) at a fixed DC field $H_{\rm DC}=0.25$~T and at a fixed AC field $H_{\rm AC} = 5$~Oe. Lower left panel: log$ (\tau $) vs log$(T_{\rm f}/T_{\rm g}-1)$ for $H_{\rm DC}=0$ and 0.25~T. The solid lines represent the fits using Eq.~(\ref{powerlawM}). Lower right panel: $ T_{\rm f} $ vs 1/ln($ \nu_{0}/ \nu $) for $H_{\rm DC}=0$ and 0.25~T. The solid lines represent the fits using Eq.~(\ref{VFequationM}).}
	\label{Fig7}
\end{figure}
For the sake of completeness, we measured AC susceptibility at different frequencies for a fixed DC field of 0.25~T and a fixed AC field of 5~Oe. As shown in the upper panel of Fig.~\ref{Fig7}, $\chi'$ manifests a peak at $T_{\rm f}$ which moves towards high temperatures and the magnitude of $\chi'$ decreases with increasing frequency. The increase in $T_{\rm f}$ with $\nu$ again supports the SG behavior of the system.
	
The relative shift in freezing temperature ($\delta T_{\rm f}$) per decade of frequency is often used as a parameter to compare different SG systems. We calculated this parameter using the relation\cite{Mydosh1993Book,Mulder1982PRB}
\begin{equation}
\delta T_{\rm{f}} = \dfrac{\Delta T_{\rm{f}}}{T_{\rm{f}}\Delta (\log \nu)},
\label{relshift}
\end{equation}
where $\Delta T_{\rm f}=(T_{\rm f})_{\nu_1}-(T_{\rm f})_{\nu_2}$ and $\Delta$log($\nu$) = log($\nu_1$) - log($\nu_2$). $\delta T_{\rm{f}}$ is also known as the Mydosh parameter.\cite{Mydosh2015RPP} For this purpose, two outermost frequencies, $\nu_1=200$~Hz and $\nu_2=10$~kHz were employed. For our system, this value is calculated to be $\delta T_{\rm f} \simeq0.017$, using the data shown in the upper panel of Fig.~\ref{Fig7}. It is about an order of magnitude larger than the values reported for canonical SG systems such as AuMn ($ \delta T_{\rm f} = 0.0045$)\cite{Mulder1982PRB} and CuMn ($ \delta T_{\rm f} = 0.005$)\cite{Mydosh1993Book} and one order of magnitude smaller than what is expected for superparamagnets (e.g. for the ideal non-interacting superparamagnetic system $ \alpha $-[Ho$ _{2} $O$ _{3} $(B$ _{2} $O$ _{3} $)], $ \delta T_{\rm f} \simeq 0.28$).\cite{Mydosh1993Book} In fact, this value is in the range usually observed for cluster SG, categorizing our system as a cluster-SG type.\cite{Mydosh1993Book,Li2001APL,Chakrabarty2014JPCM,Anand2012PRB,Maji2011JPCM,Zhang2018PRB} The value of $\delta T_{\rm f}$ essentially reflects the response or sensitivity to frequency which strongly depends on the interaction between the underlying entities. In case of magnetic clusters, the interactions between the clusters are weak and hence the sensitivity is stronger. On the other hand, in normal magnets where the interaction between magnetic ions is strong, a very large frequency is required to see any significant shift in AC susceptibility.

\begin{table*}[htbp]
	%\centering
	\setlength{\tabcolsep}{0.5 cm}
	\caption{Parameters obtained from the AC susceptibility analysis using Eqs.~\eqref{powerlawM} and \eqref{VFequationM}.}
	\begin{tabular}{c c c c c c c} 
		\hline \hline
		& & & & &  \\
		
		$ H$~(T) ~&~ $ T_{\rm g}$~(K) ~&~ $ z\nu^{\prime} $  ~&~ $\tau^{*}$~(sec) ~&~ $\tau_{0}$~(sec) ~&~ $E_{\rm a}/k_{\rm B}$~(K) ~&~ $T_0$~(K) \\
		& & & & &  \\
		\hline  
		& & & & &  \\0 & $22.0 \pm 0.1$ & $4.2 \pm 0.2$ & ($1.1 \pm 0.6) \times10^{-10}$ & ($6.6 \pm 1.5) \times10^{-9}$ & $16.0 \pm 0.6$ & $21.1 \pm 0.1$ \\
		& & & & &  \\
		0.25 & $17.8 \pm 0.1$ & $4.3 \pm 0.2$ & ($1.6 \pm 0.5) \times10^{-9}$ & ($6.2 \pm 1.6) \times10^{-8}$ & $16.4 \pm 0.6$ & $16.8 \pm 0.2$ \\
		& & & & &  \\
		\hline
	\end{tabular}
	\label{Table1}
\end{table*}
The frequency dependence of freezing temperature $T_{\rm f}$ obtained from the real part of the AC susceptibility is presented in the lower panel of Fig.~\ref{Fig7}. In SG systems, the frequency dependence of $T_{\rm f}$ can be described by the standard critical slowing down behaviour (power law), given by the dynamic scaling theory,\cite{Mydosh1993Book,Hohenberg1977RMP}
\begin{equation}
  \tau = \tau^{*}\left(\dfrac{T_{\rm f}-T_{\rm g}}{T_{\rm g}}\right)^{-z\nu^{\prime}},
  \label{powerlaw} 
\end{equation} 
where the characteristic time $\tau $ describes the dynamical fluctuation time scale and corresponds to the observation time ($t_{\rm obs} = 1/2\pi \nu$), $\tau^{*} $ is the relaxation time of a single spin flip of the fluctuating entities, $T_{\rm g}$ is the static freezing temperature as $\nu$ tends to zero, $z$ is the dynamic critical exponent, and $\nu^{\prime}$ is the critical exponent of the correlation length $\zeta = (T_{\rm f}/T_{\rm g}-1)^{-\nu^{\prime}}$. The dynamic scaling hypothesis connects $\tau$ to $\zeta$ as $\tau \sim \zeta^{z}$.
%Here, we have taken $T_{\rm g}\simeq22.2$~K and 17.8~K for $H_{\rm DC}=0$ and 0.25~T, respectively which are the values of $T_{\rm f}$ as $\nu \rightarrow 0$, obtained by extrapolating the $T_{\rm f}$ vs $\nu$ plot to lower $\nu$ values.

To fit the data, the power law in Eq.~(\ref{powerlaw}) can further be rewritten as
\begin{equation}
	\log\tau = \log\tau^{*}-z\nu^{\prime} \log\left(\dfrac{T_{\rm f}}{T_{\rm g}}-1\right).
	\label{powerlawM} 
\end{equation} 
In the lower left panel of Fig.~\ref{Fig7}, we have plotted log$(\tau)$ vs log$(T_{\rm f}/T_{\rm g} -1)$ for $H_{\rm DC} = 0$ and 0.25~T fixing $T_{g} = 22.0 \pm 0.1$~K and $17.8 \pm 0.1$~K, respectively, obtained via the best fit of the data by the power law [Eq.~(\ref{powerlaw})]. Both the curves show a linear behaviour and the obtained parameters, $ \tau^{*}$ and $ z\nu^{\prime}$ from a straight line fit [Eq.~(\ref{powerlawM})] are listed in Table~I. The dynamic scaling suggests that there is a divergence of the relaxation time at a finite transition temperature, which demonstrates a true phase transition from PM to SG in Cr$_{0.5}$Fe$_{0.5}$Ga.
The parameters, $\tau^{*}$ and $z\nu^{\prime}$ are believed to give more reliable insight into the SG dynamics. For conventional SG systems, the value of $ z\nu^{\prime} $ typically lies between $\sim 4$ and $\sim 12$ while the value of $ \tau^{*} $ ranges from $10 ^{-10} $~s to $10 ^{-13}$~s.\cite{Nam2000,Pakhira2016PRB,Ghara2014PRB,Lago2012PRB,Malinowski2011PRB}
Similarly, for the canonical SG and cluster SG, the characteristic range of $ \tau^{*}$ varies from $\sim 10^{-12} $~s to $\sim 10^{-13} $~s and $\sim 10^{-7}$ to $\sim 10^{-10}$, respectively.\cite{Lago2012PRB,Pakhira2016PRB,Mori2003PRB,Anand2012PRB,Mydosh1993Book}
Clearly, our obtained values of $ \tau^{*}$ and $ z\nu ^\prime$ fall within the ranges reported for typical cluster SG systems.\cite{Pakhira2016PRB,Lago2012PRB,Malinowski2011PRB,Mori2003PRB,Anand2012PRB} A high value of $ \tau^{*} $ also points toward the fact that in Cr$_{0.5}$Fe$_{0.5}$Ga, spin dynamics occurs in a slow manner, due to the presence of interacting clusters rather than individual spins.\cite{Anand2012PRB,Ghara2014PRB} No significant change in $z\nu^\prime$ was observed while changing the field from 0 to 0.25~T but the value of $\tau^{*}$ is changed by an order of magnitude which is still within the range expected for cluster-SG systems.

The presence of interacting clusters is also evident from the failure of Arrhenius law to fit the frequency dependent $T_{\rm f}$ data. Arrhenius law which is applicable for non-interacting or weakly interacting magnetic entities can be written as\cite{Binder1986RMP}
\begin{equation}
\tau = \tau_{0}\exp\left(\dfrac{E_{\rm a}}{k_{\rm B}T_{\rm f}}\right),
\label{Arrh}
\end{equation}
where $\tau_0$ has the same physical meaning as $\tau^*$ and $E_{\rm a}/k_{\rm B}$ is the average activation energy of the relaxation barrier. The activation energy basically measures the energy barrier in which the metastable states are separated and the Arrhenius law accounts for the time scale to overcome the energy barriers by the activation process. Our attempt to estimate $\tau_0$ and $E_a/k_{\rm B}$ from the linear fit of the ln($\tau$) vs $1/T_{\rm f}$ data in zero field yields completely unphysical values [$\tau_0 \simeq 2.1\times 10^{-62}$~s and $E_a/k_{\rm B} \simeq (3120 \pm 68)$~K]. This failure adds further support to the argument that the dynamics in our system is not simply due to single spin flips, rather it is a cooperative character due to inter-cluster interactions.

Another dynamical scaling law in spin-glass freezing is the phenomenological Vogel-Fulcher (VF) law which takes into account the interaction among the spins. According to this law, the frequency dependent $T_{\rm f}$ can be described by\cite{Mydosh1993Book,Souletie1985PRB}
\begin{equation}
\tau = \tau_{0}\exp\left[\dfrac{E_{\rm a}}{k_{\rm B}(T_{\rm f}-T_{\rm 0})}\right],
\label{VFequation}
\end{equation}
where $T_0$ is the empirical VF temperature, which is often interpreted as the interaction strength among the dynamic entities. For the purpose of fitting, it is convenient to rewrite Eq.~(\ref{VFequation}) as
\begin{equation}
\ln \tau = \ln \tau_0+\frac{E_{\rm a}/k_{\rm B}}{(T_{\rm f}-T_{0})}. 
\label{VFequationM} 
\end{equation}
Here, we show that the variation of $T_{\rm f}$ in the frequency range, which has been experimentally accessible to us, can be described by this formula. In the lower right panel of Fig.~\ref{Fig7}, the plot of $\ln \tau$ vs $1/(T_{\rm f}-T_0)$ is shown, which can be fitted well by Eq.~(\ref{VFequationM}) with $T_0 \simeq 21.1$~K and 16.8~K for $H_{\rm DC} = 0$ and 0.25~T, respectively.
The parameters, $E_{\rm a}/k_{\rm B}$ and $\tau_0$ obtained from the slope and intercept of the linear fit are summarized in Table~I. A nonzero value of $T_{0}$ and the agreement of VF law with our data suggest a finite interaction among the spins and hence the formation of clusters. The activation energy in the system is expected to be tuned under external magnetic field ($H$). It is predicted that the magnitude of the spin-glass free energy barriers ($E_{\rm a}/k_{\rm B}$) diminishes as $H^2$, the coefficient of which is proportional to the number of correlated spins.\cite{Guchhait2017PRL,Zhai2017PRB} However, our measurements at $H=0$ and $0.25$~T donot yield any visible change in $E_{\rm a}/k_{\rm B}$ which possibly suggests the role of dominant anisotropy in the spin system.
%The activation energy $E_{\rm a}/k_{\rm B}$ which also reflects the anisotropy in the system is expected to be tuned under external magnetic field.\cite{Barbara2007PRL} However, since our measurements are done only at $H=0$ and $0.25$~T which is a very small change in field, we are not able to see any visible change in $E_{\rm a}/k_{\rm B}$.

From the above assessment, it is clear that the change of relaxation time $\tau$ in our experimental frequency range can be described equally well by both power law [Eq.~(\ref{powerlawM})] and VF law [Eq.~(\ref{VFequationM})]. The obtained value of $\tau^*$ from the power law is about an order of magnitude smaller than $\tau_0$ obtained from the VF law. Such difference in characteristic time constant using two dynamical scaling laws are also reported in many cluster-SG systems e.g. Fe$_2$O$_3$,\cite{Mukadam2005PRB} Ni doped La$_{1.85}$Sr$_{0.15}$CuO$_4$,\cite{Malinowski2011PRB} etc. Nevertheless, the characteristic time constants ($\tau^*$ and $\tau_0$) obtained from both fits fall in the expected range for typical cluster-SG systems. The value of $T_{\rm g}$ is found to be larger than $T_{0}$ only a few percent, in accordance with the general trend found in the cluster-SG systems.\cite{Malinowski2011PRB}
Further, in the frame of the VF model, $T_0 \ll E_{\rm a}/k_{\rm B}$ indicates a weak coupling and $T_0 \gg E_{\rm a}/k_{\rm B}$ a strong one.\cite{Shtrikman1981PLA} For our case, $T_0$ is about $\sim 1.3E_{\rm a}/k_{\rm B}$ in zero field, which falls in the intermediate regime, suggesting a finite interaction among the magnetic entities.
%Further, the value of $T_{\rm 0}$ is found to be very close to $T_{\rm f}$ which suggests that the Ruderman-Kittel-Kasuya-Yosida (RKKY) interaction is relatively strong in our system.
Moreover, the Tholence criterion $\delta T_{\rm Th} = \frac{T_{\rm f}-T_0}{T_{\rm f}}$ is also used to compare different SG systems.\cite{Tholence1984PB} In our case, the value of $\delta T_{\rm Th}$ is calculated to be $\sim 0.06$ [taking $T_{\rm f} \simeq 22.5$~K and $T_0 \simeq 21.1$~K] at zero field which is comparable to the value reported for cluster-SG system PrRhSn$_{3}$ ($\delta T_{\rm Th} \simeq 0.076$).\cite{Anand2012PRB}

It is worth mentioning that a qualitative difference is expected between power law and VF law when the measured frequency range is large enough.\cite{Souletie1985PRB} For instance, the difference is clearly visible for Cu$_{0.954}$Mn$_{0.046}$ where the variation of $\tau$ is over 11 orders of magnitude.\cite{Souletie1985PRB} Further, closer to $T_0$ (and $T_{\rm g}$), the VF law can be adjusted to match the power law through the relation\cite{Souletie1985PRB}
\begin{equation}
ln \left(\frac{40 k_{\rm B}T_{\rm f}}{E_{\rm a}}\right) \sim \frac{25}{z\nu^\prime}. 
\label{VFPower} 
\end{equation}
Using $E_a/k_{\rm B} \simeq 16$~K, obtained from the VF law and $T_{\rm f} \simeq 22.5$~K for $\nu = 200$~Hz in Eq.~(\ref{VFPower}), the value of $z\nu^\prime$ was calculated to be $\sim 6.2$, which is slightly larger than 4.2, obtained directly from the power law fit but within the range expected for cluster-SGs.

%The expected value of $E_{\rm a}/(k_{\rm B}T_0)$ for canonical SG systems is generally close to 1 while for the cluster SG systems, it has a relatively high value. For example, $E_{\rm a}\simeq 2~k_{\rm B}T_{\rm 0}$ is reported for conventional SG system CuMn \cite{Mathieu2001PRB} and $E_{\rm a}\simeq~7~k_{\rm B}T_0$ is reported for cluster SG system La–Sr–Co–O.\cite{Fisher1988PRB} The systems Gd$_2$NiSi$_3$ and Er$_2$NiSi$_3$ with $E_{\rm a}/(k_{\rm B}T_0)$ values close to 1 and more than 1 are reported to be canonical and cluster SG systems, respectively.\cite{Pakhira2016PRB} Our estimated value of $E_{\rm a}/(k_{\rm B}T_0) \simeq 1.5$ at zero field is slightly more than 1 which is possibly due to the cluster-SG behaviour of the alloy.

\subsection{Non-equilibrium Dynamics}
\subsubsection{Magnetic Relaxations}
\begin{figure}[t!]
	\centering
	\includegraphics[width=\linewidth,height=\linewidth]{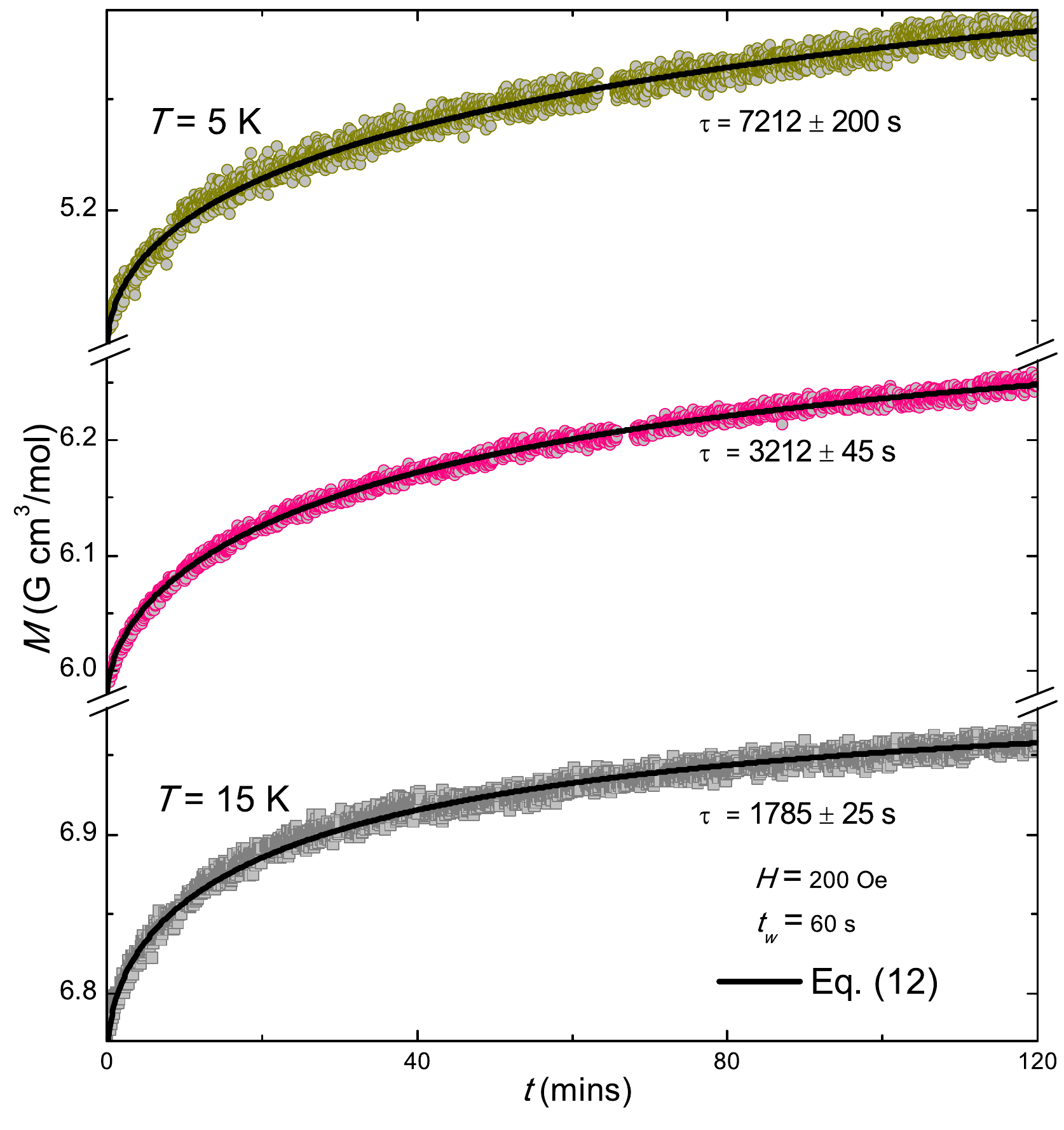}
	\caption{Relaxation of the zero-field-cooled (ZFC) magnetization measured at different temperatures $T=5$~K, 10~K, 15~K for a waiting time $t_{\rm w} = 60$~s, as discussed in the text. The solid lines represent the fit using stretched exponential function given in Eq.~\eqref{Stretched exp}.}
	\label{Fig8}
\end{figure}
Different types of glassy systems are characterized by their magnetic relaxation behaviour. To investigate such a behaviour, magnetic relaxation measurement was performed at different temperatures ($T=5$~K, 10~K, and 15~K) in the ZFC condition. The sample was cooled under zero applied field from 50~K (PM state) to the desired temperature, which is below $T_{\rm f}$. After a waiting time of $t_{\rm w}=60$~s, a magnetic field of 200~Oe was applied and the time evolution of magnetization [$M(t)$] was measured. The results are presented in Fig.~\ref{Fig8}.
%For each relaxation process, the sample was cooled from 50~K to the measured temperature 5~K in zero field. After a lapse of waiting time  $t_{\rm w}=60$, the time evaluation of magnetization $M(t)$ was measured after switching on the magnetic field of 200~Oe. The same procedure was repeated for $t_{\rm w}= 1000$~s and 5000~s.
The $M(t)$ curves follow the standard stretched exponential function
\begin{equation}
M(t)=M_{\rm 0}-M_{\rm g}\exp \left[-\left(\frac{t}{\tau}\right)^\beta \right],
\label{Stretched exp}
\end{equation}
where $M_0$ is an intrinsic magnetization, $M_{\rm g}$ is related to a glassy component of magnetization, $\tau$ is the characteristic relaxation time constant, and $\beta$ is the stretching exponent, which has values between 0 and 1 and is
a function of temperature only. Although the above function
has no specific theoretical justification, it has been widely
used to fit the magnetic relaxation data of SG systems.\cite{Chamberlin1984PRL}
In this relation, $\beta = 0$ implies that $M(t)$ is constant, i.e., no relaxation at all,
and $\beta = 1$ implies that the system relaxes with a single time
constant. Therefore, the value of $\beta$
covers the dynamics of spins with very strong to no relaxation limit. The value of $\beta$ depends on the nature of the energy barriers involved in the relaxation. For systems with a distribution of energy barriers, $\beta$ lies between 0 and 1, whereas for a uniform energy barrier, $\beta = 1$. The value of $\beta$ obtained from our fit is found to vary from 0.5 to 0.6. These values are within the range (0 to 1) of different glassy systems reported earlier.\cite{Chu1994PRL,Mydosh1993Book,Pakhira2016PRB,Chu1994PRL,Cardodo2003PRB,Khan2014PRB,Ghara2014PRB,Bhattacharyya2011PRB} Further, $\beta < 1$ signifies that the system evolves through a number of intermediate metastable states i.e. activation takes place against multiple anisotropic barriers. Moreover, the value of $\tau$ is found to increase with decreasing temperature as expected for the glassy systems, below $T_{\rm f}$.\cite{Li1998PRB,Maji2011JPCM} In fact, the values of $\tau$ obtained for Cr$_{0.5}$Fe$_{0.5}$Ga are almost comparable to that reported for other glassy systems, such as Nd$_5$Ge$_3$ (Ref.~\onlinecite{Maji2011JPCM}) and U$_2$PdSi$_3$ (Ref.~\onlinecite{Li1998PRB}).
%The value of $\tau$ is found to increase with increasing $t_{\rm w}$, indicating the stiffening of the spin relaxation effect.\cite{Pakhira2016PRB} Similarly, the relaxation measurement was also performed at different temperatures $e.g.$ $T=5$~K, 10~K, and 15~K after cooling the sample in zero-field from 50~K. At each temperature, a magnetic field of 200~Oe was applied and $M(t)$ was measured after a fixed waiting time of $t_w = 60$~s. As shown in the lower panel of Fig.~\ref{Fig7}, the data were fitted well using Eq.~\eqref{Stretched exp}. The obtained value of $\beta$ is found to vary from 0.4 to 0.6, which is again falling within the range of SG systems reported earlier.\cite{Chu1994PRL}

\subsubsection{Magnetic Memory Effect}
\begin{figure}[t!]
	\centering
	\includegraphics[width=
	\linewidth,height=1.8\linewidth]{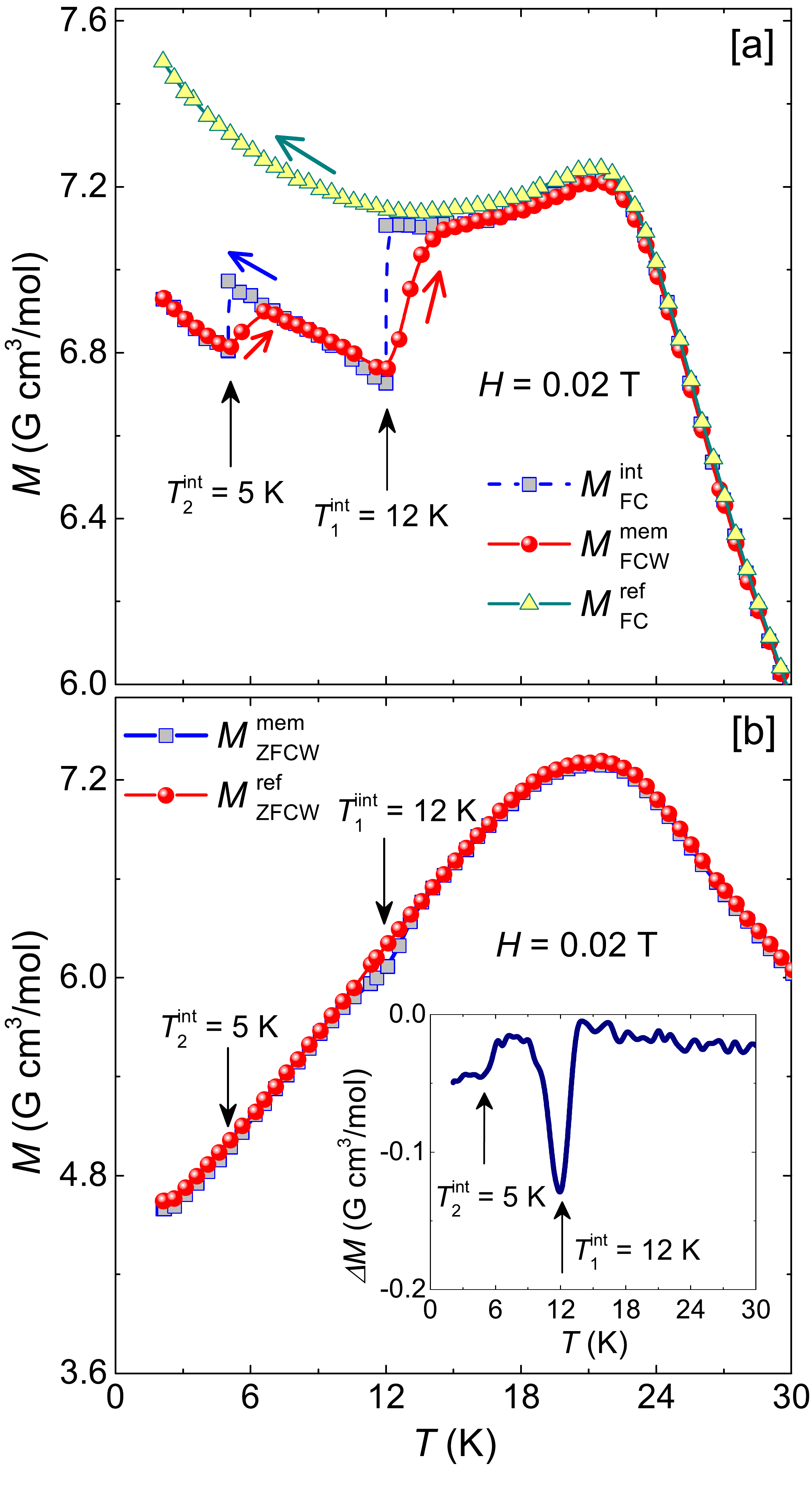}
	\caption{Memory effect as a function of temperature in (a) FC and (b) ZFC protocols in $H=200$~Oe, as discussed in the text. The measurements were interrupted at $T{^{\rm int}_{\rm 1}}=12$~K and $T{^{\rm int}_{\rm 2}}=5$~K for 2~hours each. Inset: Difference in magnetization $\Delta M=(M{^{\rm mem}_{\rm ZFCW}}-M{^{\rm ref}_{\rm ZFCW}})$ vs $T$ for the ZFC condition.}
	\label{Fig9}
\end{figure}
In order to examine the presence of non-ergodicity in the alloy and to gain new information on the low temperature dynamics, magnetic memory measurements were performed following the FC and ZFC protocols. The results are depicted in Fig.~\ref{Fig9}(a) and \ref{Fig9}(b), respectively. In the FC condition, the sample was cooled down from 50~K (PM state) to 2~K at a constant cooling rate (0.5~K/min) in an applied field of 200~Oe. The cooling process was interrupted at $T{^{\rm int}_{\rm 1}}=12$~K and $T{^{\rm int}_{\rm 2}}=5$~K for a duration of $ t_{\mathrm{w}} = 2 $~hours each. During $ t_{\mathrm{w}} $, at each temperature, the magnetic field was switched off and the system was allowed to relax. After each waiting period, the same magnetic field was switched on and the FC process was resumed. The magnetization measured during this process is denoted as $M{^{\rm int}_{\rm FC}}$ which produces steplike features at 12~K and 5~K. After reaching 2~K, the sample was heated under the same field without any interruption and $M(T)$ was recorded upto 50~K which is designated as $M{^{\rm mem}_{\rm FCW}}$. Interestingly, the obtained $M{^{\rm mem}_{\rm FCW}}$ also exhibits characteristic features at each interruption performed in $M{^{\rm int}_{\rm FC}}$, as an attempt to follow the past history of the magnetization. Thus, it is a clear signature of the magnetic memory in the system. A FC curve ($M{^{\rm ref}_{\rm FC}}$) in the same field without any interruption is also measured for reference.
%This type of FC memory effects are quite similar to those observed in various spin-glass systems.\cite{Bhattacharyya2011PRB,Jonason1998PRL} 

%In ZFC-condition as the memory effect in ZFC can only be seen in spin glass system whereas the memory effect expected to be seen in FC condition for phase-separated or superparamagnetic and spin glass or cluster glass systems. 
Similar memory effect was also measured in ZFC condition in which the sample was cooled from 50~K (PM state) to 2~K at a constant cooling rate (0.5~K/min) in zero applied field. The cooling was interrupted at $T{^{\rm int}_{\rm 1}}=12$~K and $T{^{\rm int}_{\rm 1}}=5$~K for $t_w=2$~hours each. After reaching 2~K, a magnetic field of 200~Oe was applied and $M(T)$ was recorded during warming which is designated as $M{^{\rm mem}_{\rm ZFCW}}$. For the sake of completeness, a reference curve was recorded by conventional ZFCW protocol in $H=200$~Oe which is represented as $M{^{\rm ref}_{\rm ZFCW}}$. These two curves were found to overlap with each other except at the interrupted temperature regions. This is brought out very clearly in the inset of Fig.~\ref{Fig9}(b) where the difference in magnetization $\Delta M (=M{^{\rm mem}_{\rm ZFCW}}$-$M{^{\rm ref}_{\rm ZFCW}})$ is plotted against temperature. It exhibits memory dips at each interruption points (12~K and 5~K). Thus, the observation of memory effect in both ZFC and FC conditions strengthens our assessment as cluster-SG behaviour of the compound.
%It is to be noted that the memory effect can also be detected in superparamagnetic systems in the FC condition, but not in the ZFC condition.\cite{Sasaki2005PRB,De2012JAP,Jonason1998PRL,Sun2003PRL}
%Indeed, it is only the SG systems that are known to exhibit the memory effect in both FC and ZFC conditions.\cite{Pakhira2016PRB,Bhattacharyya2011PRB} Thus, in Cr$_{0.5}$Fe$_{0.5}$Ga the appearance of memory effect below $T_{\rm f}$ during ZFC condition is an unequivocal signature of a glassy magnetic state originating from the cooperative spin-spin interactions.\cite{Sun2003PRL,Bhattacharyya2011PRB}

\begin{figure}[t!]
	\centering
	\includegraphics[width=
	\linewidth,height=\linewidth]{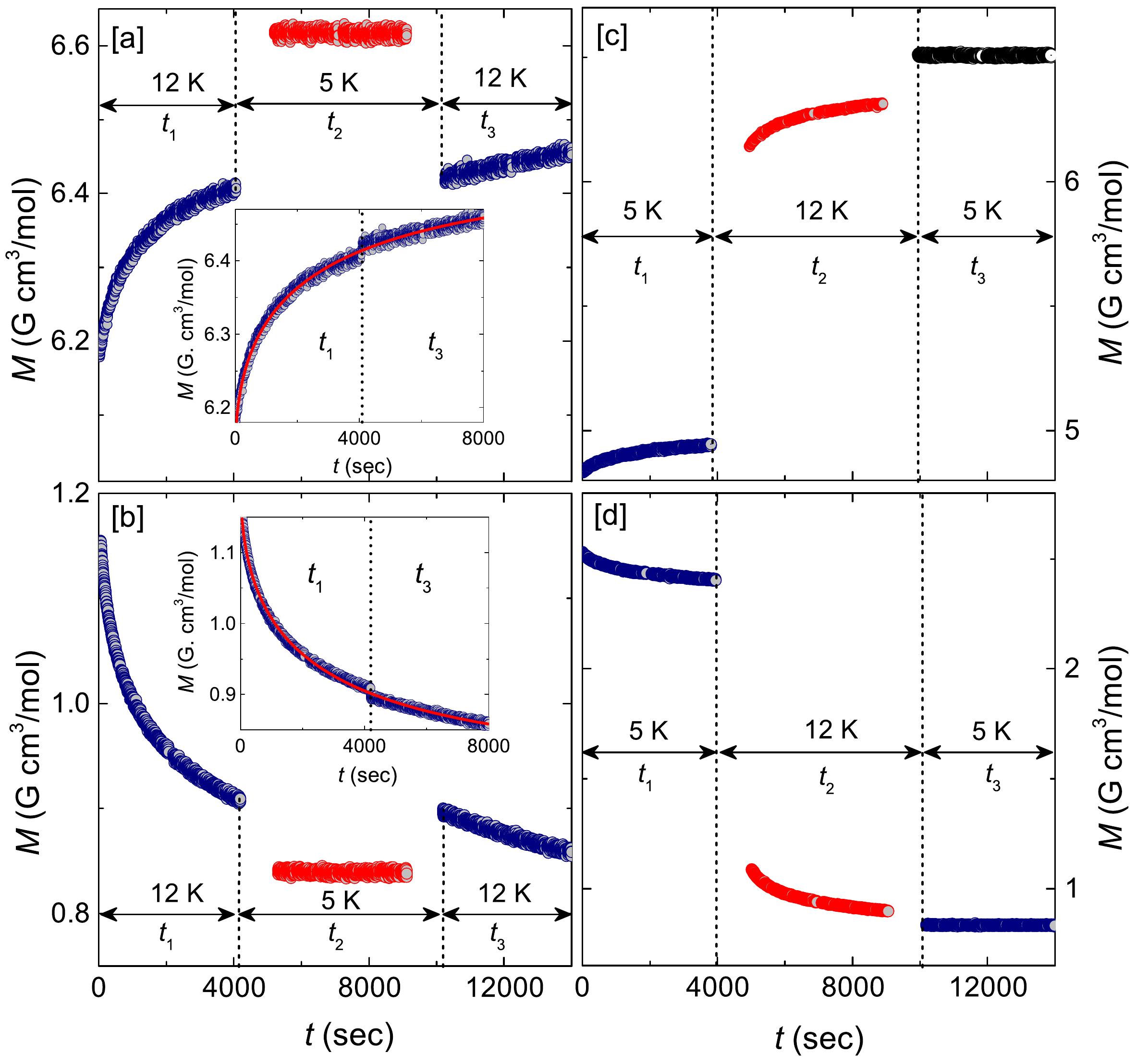}
	\caption{Magnetic relaxation measurements in the negative $T$-cycle in an applied field of $H=200$~Oe for (a) ZFC and (b) FC methods. Insets: $M(t)$ data at 12~K for negative FC and ZFC $T$-cycles along with the fit by Eq.~\eqref{Stretched exp}.
	For the positive $T$-cycle, ZFC and FC data are shown in (c) and (d), respectively.}
	\label{Fig10}
\end{figure}
To study the memory effect in further details, we performed the relaxation memory measurements for both negative and positive $T$-cycles as discussed below.   

{\em Negative $ T $-cycle :} The relaxation behaviour was recorded for the negative $T$-cycle for both ZFC and FC conditions and the results are shown in Fig.~\ref{Fig10}(a) and \ref{Fig10}(b), respectively. In the ZFC process, the sample was cooled down from 50~K to 12~K (below $ T_{\mathrm{f}} $) in zero field. At 12~K, a field of 200~Oe was applied and $M(t)$ was measured for a period of $ t_ 1=1$~hour. It is found to increase exponentially with $t$. The sample was further cooled down to 5~K in the same field and again $M(t)$ was measured for $ t_2 =1$~hour which is found to be almost constant with $t$. Subsequently, the temperature was restored back to 12~K and $M(t)$ was recorded for $ t_{\rm 3}=1$~hour in the same field which again varies exponentially with $t$. In the FC process, the sample was field cooled down to 12~K in a field of 200~Oe. At 12~K, $M(t)$ was measured for $ t_1=1$~hour after switching off the field and it was found to decay exponentially with $t$. The sample was further cooled down to 5~K in zero field and $M(t)$ was measured for $ t_2=1$~hour which is found to be constant with $t$. Finally, the sample was warmed back to 12~K in zero field and $M(t)$ was recorded again for $ t_3=1$~hour which again decays exponentially with $t$.

As shown in the insets of Fig.~\ref{Fig10}(a) and \ref{Fig10}(b), when the $M(t)$ data at 12~K measured during $t_1$ and $t_3$ put together, they simply follow a continuous growth and decay curve for the ZFC and FC processes, respectively. It indicates that the state of the sample before cooling is recovered when the sample is cycled back to the initial temperature. This is a straight forward demonstration of the memory effect in a cluster SG system where the sample remembers its previous state even after experiencing a large change in $M$. These curves were fitted by the stretched exponential function [Eq.~\eqref{Stretched exp}] with $\beta \simeq 0.5$, similar to that observed in the magnetic relaxation measurements.
%This type of behavior is also an indication that the memory effect is quite strong in this alloy, below $T_{\rm f}$.

{\em Positive $ T $-cycle:} Similar to the negative $T$-cycle, both ZFC and FC relaxation behaviours were also recorded for the positive $T$-cycle and are shown in Fig.~\ref{Fig10}(c) and \ref{Fig10}(d), respectively. In the ZFC process, the sample was cooled down from 50~K to 5~K in zero field. At 5~K, a field of 200~Oe was applied and $M(t)$ was recorded for $ t_{1}=1$~hour which shows a gradual increase with $t$. The sample was then heated upto 12~K in the same field and again $M(t)$ was measured for $ t_{2}=1$~hour which also shows a gradual increase with $t$. Finally, the temperature was restored back to 5~K but the $M(t)$ measured for another $ t_{3}=1$~hour is found to be $t$ independent. In the FC process, the sample was cooled down to 5~K in a field of 200~Oe. At 5~K, magnetic field was switched off and the same sequence (as for the ZFC process) was repeated. As shown Fig.~\ref{Fig10}(d), the obtained results follow the same trend as for the ZFC sequence but in the opposite direction. It is evident that unlike the negative $T$-cycle, there is no continuity in the $M(t)$ data measured during $t_1$ and $t_3$ at 5~K suggesting that the nature of the magnetic relaxation during $t_3$ is quite different from that during $t_1$. Thus, positive $T$-cycling revives the magnetic relaxation process and no magnetic memory effect is observed when the temperature is restored.

The memory effect in SG systems has been widely studied via magnetization measurements. This phenomena is usually discussed in the framework of two theoretical models: the droplet model\cite{Fisher1988PRB} and the hierarchical model.\cite{Lefloch1992EPL} These are two well established models which are successfully applied in several experimental studies.\cite{Pakhira2016PRB,Sun2003PRL} At a given temperature, a multi-valley spin structure is organized on the free-energy landscape in the hierarchical model, whereas in the droplet model only one spin configuration is favoured. Basically, in the hierarchical model, these free energy valleys which are metastable states split into new sub-valleys as the temperature is lowered and get merged with increasing temperature. This picture obviously give rise to the observed memory effects. When the temperature of the system is lowered from $T$ to $T-\Delta T$, each valley splits into a set of sub-valleys. If $\Delta T$ is large, the energy barriers separating the main valleys become too high and the system cannot overcome these barriers during the waiting time $t_2$. Therefore, the relaxation occurs only within the sub-valleys of each set. As the temperature is brought back to $T$, the sub-valleys and barriers merge back to the original free-energy landscape and the relaxation at $T$ is not at all disturbed by the intermediate relaxations at $T-\Delta T$. But when the temperature of the system is increased from $T$ to $T+\Delta T$, the barriers between the free energy multi valleys are lowered or even get merged. Therefore, the relaxations can occur within different valleys. When the temperature is lowered back to $T$, although the free-energy landscape is restored, the relative occupancy of each valley does not remain the same as before. Thus, the state of the system changes after a temporary heating cycle showing no memory effect.

Experimentally, these two models can be distinguished by studying the influence of $T$-cycling on magnetic relaxation.  
In the droplet model, the original spin configuration is restored after a $T$ cycling  $i.e.$ one would expect a symmetric behaviour in magnetic relaxation with respect to the positive/negative $T$-cycling.
On the other hand, in the hierarchical model, the original spin configuration is destroyed after a positive $T$-cycling and one would expect an asymmetric response (or, no memory effect) in magnetic relaxation. Thus, based on the above criteria, our observed asymmetric response in the positive $T$-cycle during both ZFC and FC processes supports the hierarchical organization of the metastable states in the cluster-SG system. Since the hierarchical organization requires a large number of degrees of freedom to be coupled, it can not be produced simply by the independent behaviour of individual spins and consequently highlights the important role played by inter-particle/intra-cluster interactions.

\section{Summary}
In summary, we present a detailed and a systematic study of the structural and magnetic properties of Cr$_{0.5}$Fe$_{0.5}$Ga. No evidence of any structural disorder was found from the temperature dependent powder XRD measurements down to 15~K. The temperature dependent DC magnetization shows the onset of a SG transition at low temperatures which is caused by magnetic site disorder and magnetic frustration due to competing AFM and FM interactions. The SG transition is further justified by AC susceptibility measurements. The results clearly indicate that the fitted parameters, as obtained from the relative shift in $T_{\rm f}$ and the dynamical scaling laws, are consistent with that expected for cluster SG systems. The activation energy of the metastable states is estimated to be $E_{\rm a}/k_{\rm B} \simeq 16$~K. A clear signature of the magnetic memory effect was observed below the freezing temperature in both FC and ZFC processes further demonstrating the cluster-SG behavior of the compound under investigation. In the positive $T$-cycle, a small heating reinitializes the relaxation process and the magnetization is unable to restore its initial value. Such an asymmetric response of magnetic relaxation with respect to positive temperature change favours the hierarchical model. The Debye temperature estimated from the low temperature $C_{\rm p} (T)$ data is consistent with that obtained from the $V_{\rm cell} (T)$ analysis. Although our experimental results point towards the formation of cluster SG state, the underlying mechanism behind such a formation is not yet understood. Further studies preferably neutron scattering and $\mu$SR experiments may provide useful insite.

\section{Acknowledgement}
We would like to acknowledge BRNS, India for financial support bearing sanction No.37(3)/14/26/2017-BRNS. We thanks B. R. Sekhar for his valuable suggestions. PB is supported by the IISER-TVM post-doctoral fellowship programme.

%\bibliography{BIBSUM}

\providecommand{\noopsort}[1]{}\providecommand{\singleletter}[1]{#1}%

\end{document}